\documentclass[]{aa} 
\usepackage{graphicx}
\usepackage{natbib}
\usepackage{epsfig}

\usepackage{txfonts}
\newcommand{\kep}{{\sl Kepler}}
\newcommand{\mass}{{2-MASS}}

\def\ms{\,m\,s$^{-1}$}         
\def\kms{\hbox{\,km\,s$^{-1}$}}       
\def\vsini{\hbox{$v$\,sin\,$i$}}      
\def\Ms{\hbox{$M_{\star}$}}             
\newcommand{\Msun}{$M_{\odot}$}             
\def\Rsun{\hbox{$R_{\odot}$}}
\def\Rs{\hbox{$R_{\star}$}}

\def \Mp{$M_{\rm p}$}
\def \Rp{$R_{\rm p}$}
\newcommand{\vrad}{$v_{\rm rad}$} 
\newcommand{\teff}{T$_{\rm eff}$}
\newcommand{\logg}{log {\it g}}
\newcommand{\vmic}{v$_{mic}$}
\newcommand{\vmac}{v$_{macro}$}

\newcommand{\feh}{[Fe/H]}
\newcommand{\mtier}{$M_\star^{1/3}/R_\star$} 
\newcommand{\MJ}{{\it M}$_{\rm Jup}$}
\newcommand{\ME}{{\it M}$_\oplus$}
\newcommand{\RJ}{{\it R}$_{\rm Jup}$}
\newcommand{\sn}{S/N}

\begin{document}
   \title{SOPHIE velocimetry of Kepler transit candidates \\
   XI. Kepler-412 system: probing the properties of a new inflated hot Jupiter}

\author{
M. Deleuil\inst{\ref{lam}}
\and J.-M. Almenara\inst{\ref{lam}} 
\and A. Santerne\inst{\ref{lam},\ref{caup}}
\and S.C.C. Barros\inst{\ref{lam}} 
\and M. Havel\inst{\ref{oak}} 
\and G. H\'ebrard\inst{\ref{ohp},\ref{iap}}
\and A.S. Bonomo\inst{\ref{inaf}}
\and F. Bouchy\inst{\ref{lam}}
\and G. Bruno\inst{\ref{lam}}
\and C. Damiani \inst{\ref{lam}}
\and R.F. D\'iaz\inst{\ref{lam}}
\and G. Montagnier\inst{\ref{ohp},\ref{iap}}
\and C. Moutou\inst{\ref{lam}} 
          }

\institute{
Aix Marseille Universit\'e, CNRS, LAM (Laboratoire d'Astrophysique de Marseille) UMR 7326, 13388, Marseille, France \\\email{magali.deleuil@lam.fr} \label{lam}
\and Centro de Astrof\'isica, Universidade do Porto, Rua das Estrelas, 4150-762 Porto, Portugal\label{caup}
\and NASA Ames Research Center, Moffett Field, CA 94035\label{oak}
\and Observatoire de Haute Provence, 04670 Saint Michel l'Observatoire, France\label{ohp}
\and Institut d'Astrophysique de Paris, 98bis boulevard Arago, 75014 Paris, France\label{iap}
\and INAF Osservatorio Astrofisico di Torino, Via Osservatorio 20, 10025 Pino Torinese, Italy\label{inaf}
             }

   \date{Received 08 November 2013 / Accepted 12 December 2013 }

 
  \abstract
   {Hot Jupiters are still a fascinating exoplanet population that presents a diversity we are still far from understanding. High-precision photometric observations combined with radial velocity measurements give us a unique opportunity to constrain their properties better, on both their internal structure and their atmospheric bulk properties.}
   {We initiated a follow-up program of \kep-released planet candidates with the goal of confirming the planetary nature of a number of them through radial velocity measurements. For those that successfully passed the radial velocity screening, we furthermore performed a detailed exploration of their properties to characterize the systems. As a byproduct, these systematic observations allow us to  consolidate the exoplanets' occurrence rate.  }
   {We performed a complete analysis of the Kepler-412 system, listed as planet candidate KOI-202 in the \kep\ catalog, by combining the \kep\ observations from Q1 to Q15, to ground-based spectroscopic observations that allowed us to derive radial velocity measurements, together with the host star parameters and properties. We also analyzed the light curve to derive the star's rotation period and the phase function of the planet, including the secondary eclipse. 
    }
   {We secured the planetary nature of Kepler-412b. We found the planet has a mass of 0.939 $\pm$ 0.085 \MJ\ and a radius of 1.325 $\pm$ 0.043 \RJ\ which makes it a member of the bloated giant subgroup. It orbits its G3 V host star in 1.72 days. The system has an isochronal age of 5.1~Gyr, consistent with its moderate stellar activity as observed in the \kep\ light curve and the rotation of the star of 17.2 $\pm$ 1.6 days.  From the detected secondary we derived the day side temperature as a function of the geometric albedo. We estimated that the geometrical albedo  A$_g$ should be between 0.094 $\pm$ 0.015 and 0.013  $^{+0.017}_{-0.013}$ and the brightness of the day side 2380 $\pm$ 40~K. The measured night side flux corresponds to a night side brightness temperature of 2154 $\pm$ 83~K, much greater than what is expected for a planet with homogeneous heat redistribution. From the comparison to star and planet evolution models, we found that dissipation should operate in the deep interior of the planet. This modeling also shows that despite its inflated radius, the planet presents a noticeable amount of heavy elements, which accounts for a mass fraction of 0.11 $\pm$ 0.04.}
   {}

   \keywords{stars: planetary systems - stars: fundamental parameters - techniques: photometry - techniques:
  radial velocities - techniques: spectroscopy - stars : individual : Kepler-412  }

\titlerunning{Kepler-412b : a new inflated hot Jupiter with a low albedo}
\authorrunning{M. Deleuil}

   \maketitle
%

\section{Introduction}
Space-based photometry observations of transiting planets have opened a new area for characterizing planets. Combined with high-precision radial velocity measurements, photometric observations allow deriving accurate planet parameters. In addition to a complete determination of the orbital parameters, it indeed provides valuable hints of the interior structure of the planet that might be related to the formation process and evolution history of the system. For the close-in giant planets, this has allowed the large diversity to be quantified better among what might appear as a single family at first glance. While most of these massive planets seem to possess a core whose typical mass is estimated at about 10\ME\ \citep{Guillot2006,Miller2011,Fortney2011K12}, some such as Kepler-7b \citep{Latham2010} appear to lack such a core, whereas some such as  CoRoT-20b \citep{Deleuil2012} present a surprisingly large amount of heavy material in their interior that could account for a core of a few hundreds Earth masses.

   \begin{figure*}[ht]
   \centering
   \includegraphics[height=11cm,width=17cm]{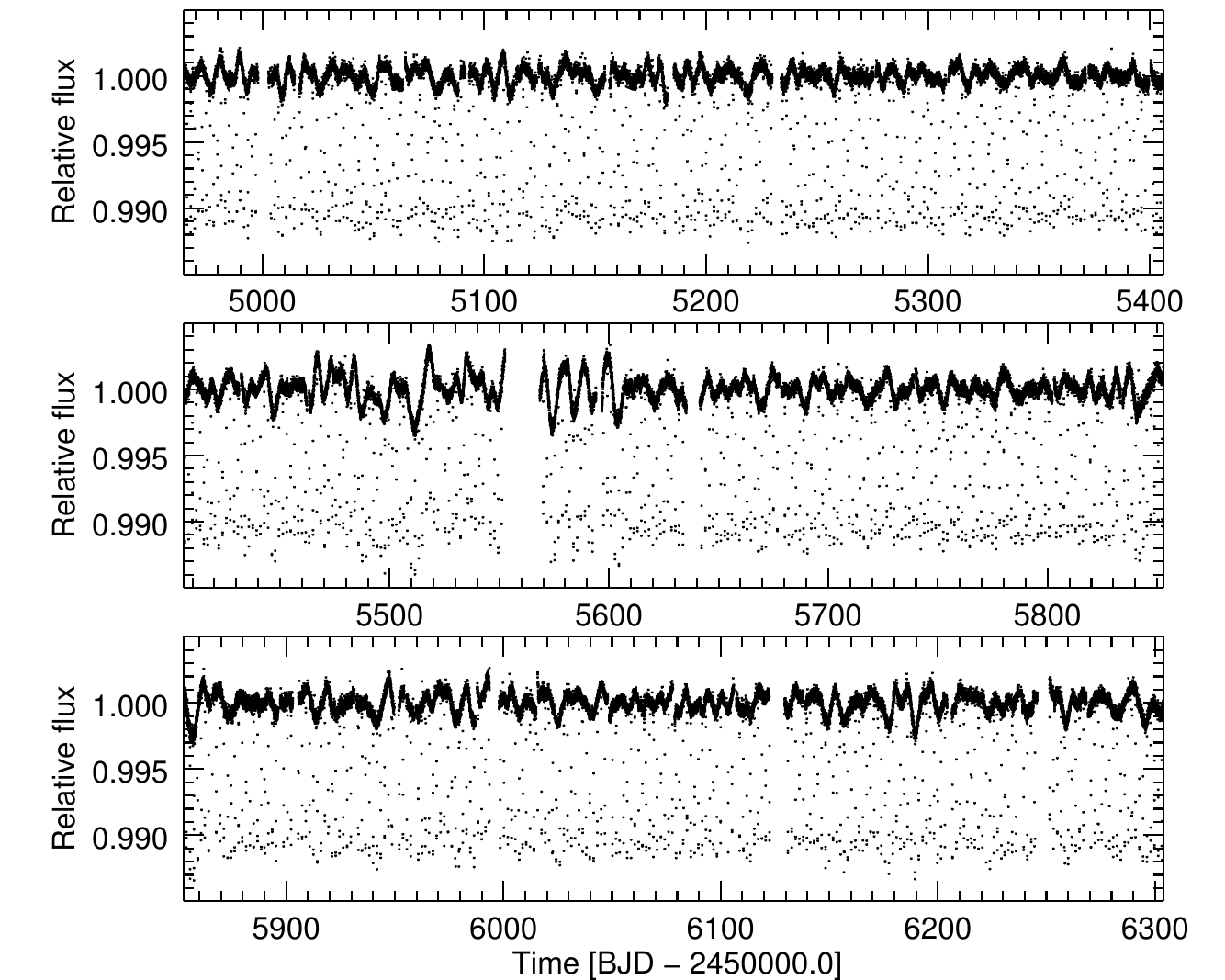}
   \caption{Q1 to Q15 photometric times series of Kepler-412 observations reduced by the \kep\ pipeline and normalized. }
              \label{LC}
    \end{figure*}

Probing the atmospheric properties of exoplanets is another situation that still remains challenging owing to the small amplitude of the planet's atmospheric signatures. Significant advances have been, however, achieved during the past ten years from space-based observations in the near- or mid-infrared thanks to {\sl Spitzer} observations \citep[e.g.]{Grillmair2007} or in some rare cases, from ground-based observations \citep{CollierCameron2002,Berdyugina2011}. While the infrared spectral range gives access to the thermal emission, at visible wavelength, in principle, one can measure the reflected light of the atmosphere over the planet orbit and further obtain some hints on the energy budget of the planet. CoRoT and, even more, \kep\ have proven such measurements are from now on possible for hot Jupiters, such as  CoRoT-1 \citep{Snellen2009,Alonso2009b} or Kepler-17b \citep{Desert2011}, but also Super-Earth ones \citep{SanchisOjeda2013}, thanks to the long temporal coverage and the very high photometric precision that could be achieved from space-based observations. However, the number of these planets for which a noticeable geometric albedo has been measured remains limited today, confirming that hot Jupiters have very low reflectivity in visible, with an upper atmosphere likely lacking the bright clouds that are present in the atmosphere of the much colder giant planets in our solar system.

In this paper we confirm the planetary nature of one planet candidate in the \kep\ catalog \citep{Borucki2011ApJ728} thanks to radial velocity observations and the planet's subsequent characterization. This KOI, whose ID is KOI-202, is indeed one of the Kepler targets in the giant domain we observed during our radial velocity campaigns with the SOPHIE \citep{Bouchy2011,Santerne2012Fp} and ESPaDOnS spectrographs. It was identified as a potentially valuable target since a rough analysis of the \kep\ light curve shows a clear secondary. This seemed a good opportunity to add a new member to the handful of giants for which the atmospheric properties could be explored.

Section 2 of this paper presents the complete set of observations that we used to assess the planetary nature of Kepler-412b and measure its parameters, that is, the complete sequences of \kep\ observation and our follow-up ground-based spectroscopic observations. The combined analysis of both sets of data that allowed us to determine the system parameters is described in Section 3. The analysis of the phase curve and the secondary with the estimates of the planet's albedo, redistribution factor, and brightness temperatures are presented in Section 4. The constraints we achieved on the planet's internal structure are discussed in Section 5 and those on its atmospheric properties in Section 6. 
Section 7 summarizes our findings for this new system. 

\begin{table}[h]
\caption{ Kepler-412 IDs, coordinates, and magnitudes.}       
\label{ID}     
\begin{minipage}[!]{7.0cm}  
\renewcommand{\footnoterule}{}     
\begin{tabular}{ccc}       
\hline\hline                 
Kepler ID & 7877496 \\
KOI ID & 202 \\
USNO-B1 ID  & 1336-00316815 \\
2MASS ID   &  19042647+4340514  \\
GSC2.3 ID & N2EO000653  \\
\\
\multicolumn{2}{l}{Coordinates} \\
\hline            
RA (J2000)  & 19:04:26.48  \\
Dec (J2000) & 43:40:51.46 \\
\\
\multicolumn{3}{l}{Magnitudes} \\
\hline
Filter & Mag & Error \\
\hline
K$_p^a$ &   14.31 & \\
V$^b$ & 13.73 & 0.24 \\
g$^a$ & 14.727 & 0.04 \\ 
r$^a$ & 14.240  & 0.04 \\ 
i$^a$ & 14.130  & 0.04 \\ 
z$^a$ & 14.077  & 0.04 \\ 
J$^c$  & 13.190 & 0.027 \\
H$^c$  & 12.932 & 0.033 \\
K$^c$  &  12.837 &  0.031\\
W1$^d$  &  12.824 &  0.028\\
W2$^d$  &  12.860 &  0.031\\
\hline
\vspace{-0.5cm}
\footnotetext[1]{from the \kep\ Input Catalog;}
\footnotetext[2]{from the second-generation guide star catalog \citep{Lasker2008};}
\footnotetext[3]{from the \mass\ catalog \citep{Skrutskie2006}.}
\footnotetext[4]{from the WISE catalog \citep{Cutri2012}.}
\end{tabular}
\end{minipage}
\end{table}

\section{Observations and data reduction }
\subsection{Kepler photometric observations}
\label{photom}
Kepler-412b is one of the planet candidate listed in the \kep\ data release of Feb. 2, 2011. From the data obtained in Q0 to Q5, \cite{Borucki2011ApJ728} indeed report the detection of single periodic transits in the light curve of the star with a period of 1.72 days, a depth of 1\%, and a duration of 2.1 hour, making this candidate a potential member of the giant-planet population. Table~\ref{ID} gives the various IDs, coordinates and magnitudes of the host star as could be found from different catalogs.

For the study presented in this paper, we used the complete sequence of observations collected by the \kep\ spacecraft during its more than four years of operation, that is, between May 13, 2009 and August 4, 2013  (Q1 to Q15). Most of the observations were performed in long cadence mode (temporal sampling of 29.4 min) but three quarters during which data were acquired in high cadence mode. The light curve of quarters Q1 to Q15 were downloaded from the Mikulski Archive at the Space Telescope Institute. These time series are reduced by the Photometric Analysis Kepler pipeline (version 8.01), which includes corrections for the main instrumental effects: background subtraction, barycentric, cosmic ray, and мargabrighteningМ effect \citep{Jenkins2010}.  Figure~\ref{LC} displays the \kep\ observations where the quarters were normalized by the median value of the flux and gathered together. The transits are the most prominent features, but the photometric series shows a modulation of a few ppm related to a moderate stellar activity but with a sudden increase starting around BJD 2455000 and ending up at BJD 2456000. 

Taking advantage of the stellar spot imprints in the light curve, we seek the rotational period of the star. 
The transits were removed from the light curve by subtracting a transit model derived from the full series of transits. A trend was then subtracted from this light curve corrected from transits, and residuals at the planet period were calculated from the light curve binned at the transit period. These binned residuals at the phase of the planet orbit were subtracted from the original light curve, once interpolated and set on the temporal scale. The goal of this procedure was to remove any signal related to the planet in the data as properly as possible and keep the star's activity signature only in the light curve. Finally, outliers were corrected thanks to a three-sigma clipping. 

   \begin{figure}
   \centering
   \includegraphics[height=7cm,width=9cm]{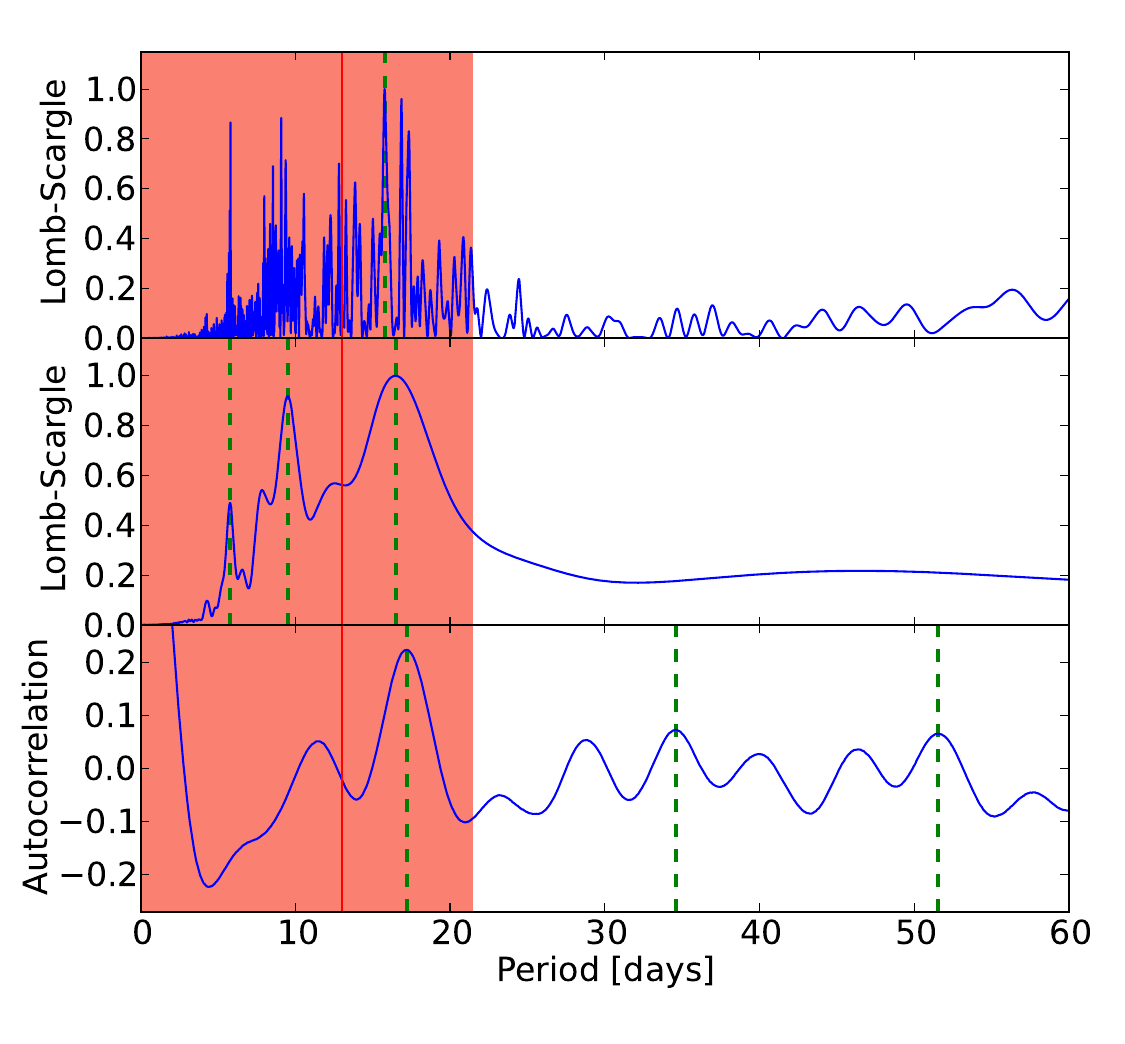}
  \caption{{\sl Top panel:}  Lomb-Scargle periodogram of the complete light curve.  {\sl Middle panel:} Lomb-Scargle periodogram obtained  from 67-day chunks of the light curve. {\sl Bottom panel:} light curve autocorrelation.  The green dotted lines indicate the center of  Gaussians fitted to the data to derive the main frequency and its harmonics. The red vertical line corresponds to the rotation period deduced from the \vsini\ and the star's radius, assuming the star's rotation axis is perpendicular to the line of sight, and the shaded areas are the associated 1$\sigma$, 2$\sigma$, and 3$\sigma$ upper limit regions from the darkest to the lightest one.}
              \label{LombScargle}
    \end{figure}

The Lomb-Scargle periodogram of the complete light curve cleaned from the signal at planetary period does not give a clear result but a forest of peaks (Fig.~\ref{LombScargle}) with no clearly dominant one. This is likely due to the evolution of the active regions and their area along successive rotations that are very short compared to the four-year time span of the \kep\ observations.  
To minimize the possible effects of surface features evolution, we calculated the Lomb-Scargle periodograms on the light curve split in 67-day time spans, that is 20 chunks with the same duration. This time span appeared to be long enough to provide an appropriate coverage of the surface phenomenon but short enough to weaken long-term surface effects. The resulting mean periodogram displays three significant peaks at 16.5 $\pm$ 2.6 days, 9.51 $\pm$ 0.71 days, and 5.78$\pm$ 0.26 days (Fig.~\ref{LombScargle}). We conclude that these frequencies correspond to the stellar rotation period with its first two harmonics at half and one third of this frequency, respectively.

As an additional verification, we also computed the light curve autocorrelation \citep{McQuillan2013}. We found a significant peak at 17.2 $\pm$ 1.6 days as shown in the bottom panel of Fig.~\ref{LombScargle}. The two first harmonics at  34.6 $\pm$ 1.4 days and 51.5 $\pm$ 1.6 are also detected but do not clearly prevail. We noticed that this 17.2-day peak is ten time the period of the transit. Even though the transits were properly removed by using a model, we cannot completely exclude this peak being due to some residuals at the transit period. We checked how these values compared to the rotation period that is deduced from the projected rotation velocity \vsini\ (see Sect.~\ref{spectro}). Assuming the rotation angle of the star is perpendicular to the line-of-sight ,we derived a star's rotation of 13.0 $\pm$ 2.8 days from the  \vsini\ of 5.0 $\pm$ 1.0 \kms. As highlighted in Fig.~\ref{LombScargle}, this value  is indeed compatible with the values we derived from the light curve analysis. We finally adopted the value of 17.2 $\pm$ 1.6 days, which provides the smallest uncertainty.

\subsection{Spectroscopic observations}
\label{spectro}
Kepler-412 was observed with the SOPHIE spectrograph on the 1.93-m telescope at the Observatoire de Haute Provence (France) as part of our Kepler candidates follow-up program \citep{Bouchy2011,Santerne2012Fp}. The observations were performed in high-efficiency mode without acquiring the simultaneous thorium-argon calibration, because we used the second fiber to monitor the Moon background light. A first set of spectra were acquired on July 2011 (see Table~\ref{TabRV}) and confirmed the planetary nature of Kepler-412b. The radial velocity measurements were in phase with the transit period with a semi-amplitude consistent with what is expected for a massive planet. In addition, we found no evidence of a close-by star, whether in the background or gravitationally bound to our target from the line-bisector analysis of the CCFs or from the use of cross-correlation masks constructed for different spectral types  \citep{Bouchy2008}. These measurements were completed seven months later when the Kepler field became visible again from the ground. The complete RV measurements gives a semi-amplitude close to 140 \ms which corresponds to a Jupiter-mass companion.

\begin{table}[ht]
\begin{center}{
\caption{\label {TabRV}  Log of SOPHIE radial velocity observations}
\begin{tabular}{lllll}
\hline
\hline
Date  & HJD    & \vrad\  & $\sigma$\vrad  & \sn  \\
          &           & \kms    &  \kms           &  550~nm \\ 
\hline
\hline
2011-07-25 & 2455767.53019 & -54.302  &   0.018 & 15  \\
2011-07-31 & 2455773.49310 & -54.576  &  0.010  & 18 \\
2011-07-31 & 2455774.43481 & -54.307   & 0.012 & 18 \\
2012-02-26 & 2455983.66197 & -54.530   & 0.042  & 10 \\
2012-03-24 & 2456010.62956 & -54.474   & 0.014  & 14\\
2012-03-26 & 2456012.61452 & -54.597   & 0.020 &  15 \\
\hline
\end{tabular}}
\end{center}
\end{table}

Once the planetary nature of Kepler-412b was confirmed, we observed the star with ESPaDOnS at the 3.6-m Canada-France-Hawaii Telescope in Mauna Kea as part of the 12AF95 program. The spectrograph provides spectral coverage of the 370 - 1000 nm region at a spectral resolution of 65 000. The star was observed  in `object+sky' mode on July 2, 2012,  in a series of short exposures totalizing 15520 sec. We used the individual spectra  reduced by the CFHT Upena/Libre-Esprit pipeline and co-added them once they were set in the rest frame. This results in a spectrum with a \sn\ of 110  in the continuum per resolution element at 6000~\AA.

The spectral analysis was carried out with the semi-automatic software package VWA \citep{Bruntt2010MNRAS,Bruntt2010AA}, which derives the atmospheric parameters by minimizing the correlations of the \ion{Fe}{i} abundance with both equivalent width and excitation potential. We used the  \vsini\ derived from the SOPHIE CCFs and later checked its value and the \vmac\ one on a set of isolated spectral lines. An estimate the surface gravity was derived from the \ion{Fe}{i}  and \ion{Fe}{ii} agreement, but we also carried out an independent estimate  from the pressure-sensitive lines: the \ion{Mg}{1}b lines and the \ion{Ca}{i} ones at 6122\AA\ and 6262\AA. The abundances of the elements are given in Table~\ref{abund} from metals with more than ten lines with measured abundances. Taking uncertainty on \teff, \logg, and \vmic\ into account, we calculated a mean metallicity of [M/H] = 0.27 $\pm$ 0.12.  The final atmospheric parameters of the host star are given in Table~\ref{AllParams}.

To determine the star's fundamental parameters: mass, radius, and age, we compared the atmospheric parameters \teff\ and \feh\ and the stellar density from the transit fit to a two-dimensional grid of STAREVOL evolutionary tracks of stellar mass and metallicity \cite[Palacios, {\sl priv. com.};][]{Lagarde2012}.  We calculate the probability distribution function by minimizing the distance to the grid of evolutionary tracks from the Monte Carlo realizations of the atmospheric parameters and their associated errors and the distribution of stellar density provided by the transit modeling (see Sect.~\ref{modeling}).  This yields  \Ms\ = 1.15 $\pm$ 0.06 \Msun, and \Rs\ = 1.31 $\pm$ 0.03 \Rsun. These values are consistent with those obtained from the complete data set modeling performed with PASTIS and given in  Table~\ref{AllParams}. 

\begin{table}
 \centering
 \caption{Relative abundances of the main elements
 \label{abund}}
 \begin{footnotesize}
\begin{tabular}{llr}
\hline
\hline
  Element        & Abundances & Nb lines \\
\hline
  {Li  \sc   i} &     0.98  $\pm$ 0.10  &   3   \\ 
  {O  \sc   i} &     0.26  $\pm$ 0.10  &   3   \\ 
  {Na  \sc   i} &     0.48  $\pm$ 0.10  &   2   \\ 
  {Si \sc   i} &     0.28  $\pm$ 0.08   &  13   \\ 
  {Ca \sc   i} &    0.27 $\pm$ 0.09   &   5   \\ 
  {Ti \sc   i} &     0.22  $\pm$ 0.09  &  14  \\ 
  {Ti \sc  ii} &     0.44  $\pm$ 0.08   &   6   \\ 
  {Cr \sc   i} &     0.19  $\pm$ 0.08   &   5   \\ 
  {Mn \sc   i} &     0.43 $\pm$ 0.10    &   5  \\ 
  {Fe \sc   i} &     0.26  $\pm$ 0.08   & 104   \\ 
  {Fe \sc  ii} &     0.29  $\pm$ 0.09   &  16   \\ 
  {Co \sc   i} &     0.42  $\pm$ 0.09  &   5   \\ 
  {Ni \sc   i} &     0.32 $\pm$ 0.08   &  34  \\ 
\hline
\end{tabular}
\end{footnotesize}
\end{table}

With an age of 5.1 $\pm$ 1.2 Gyr, the star is well on the main-sequence phase, which makes it a mature system. The moderate star's rotation period we derived from the light curve analysis also agrees with such an evolutionary status. Inspection of the spectra reveals no emission in the \ion{Ca}{ii} H and K lines or other photospheric lines but a detectable \ion{Li}{i} doublet line. According to \cite{Sestito2005}, the measured \ion{Li}{i} abundance of 2.03, points toward an age for the star $>$ 1~ Gyr.  All the age indicators, including the moderate photometric variation observed in the light curve of the host star, agree and confirm the inferred stellar age.

\section{Kepler-412 system analysis}
\label{modeling}
The properties of the Kepler-412 system were derived from a joint analysis of the light curve, the radial velocity measurements, and the star's properties. This analysis was carried out with the Planet Analysis and Small Transit Investigation Software (PASTIS)  \citep{Diaz2013b}. Primarily developed for planet validation \citep{MoutouC22}, the software models all the observables from the light curve, the spectral energy distribution and radial velocity measurements in a self-consistent way and further provides rigorous estimate of the confidence regions for all parameters \citep[see][e.g.]{Diaz2013KOI205}.  
 
   \begin{figure}
   \centering
   \includegraphics[height=6cm,width=9cm]{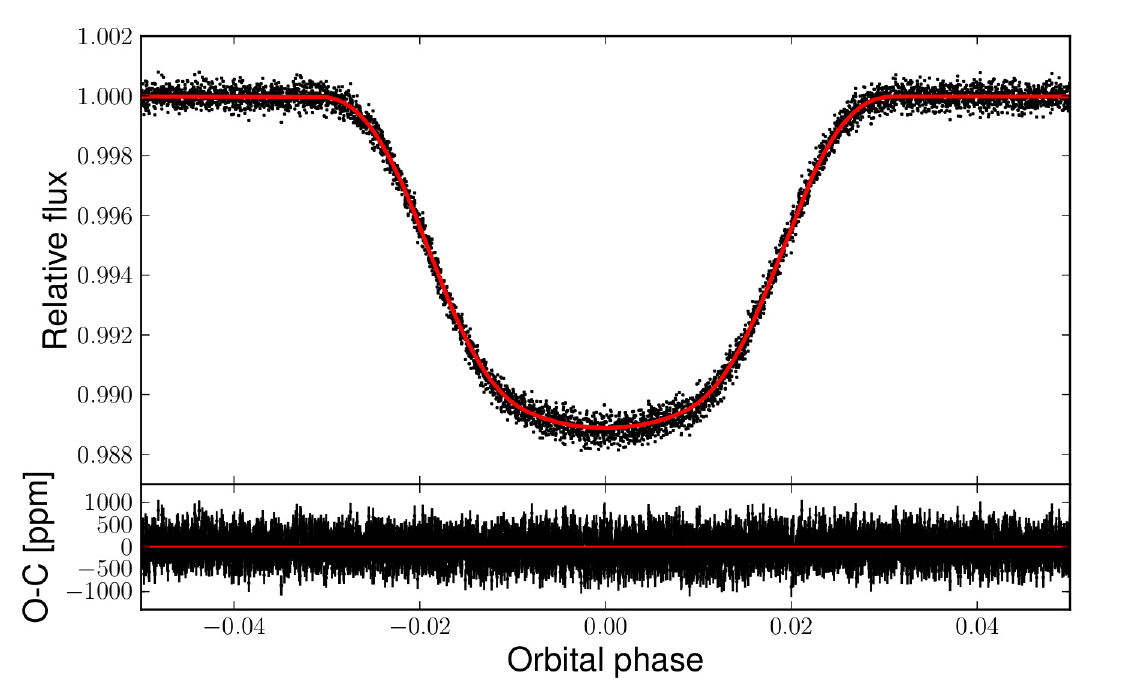}
   \includegraphics[height=6cm,width=9cm]{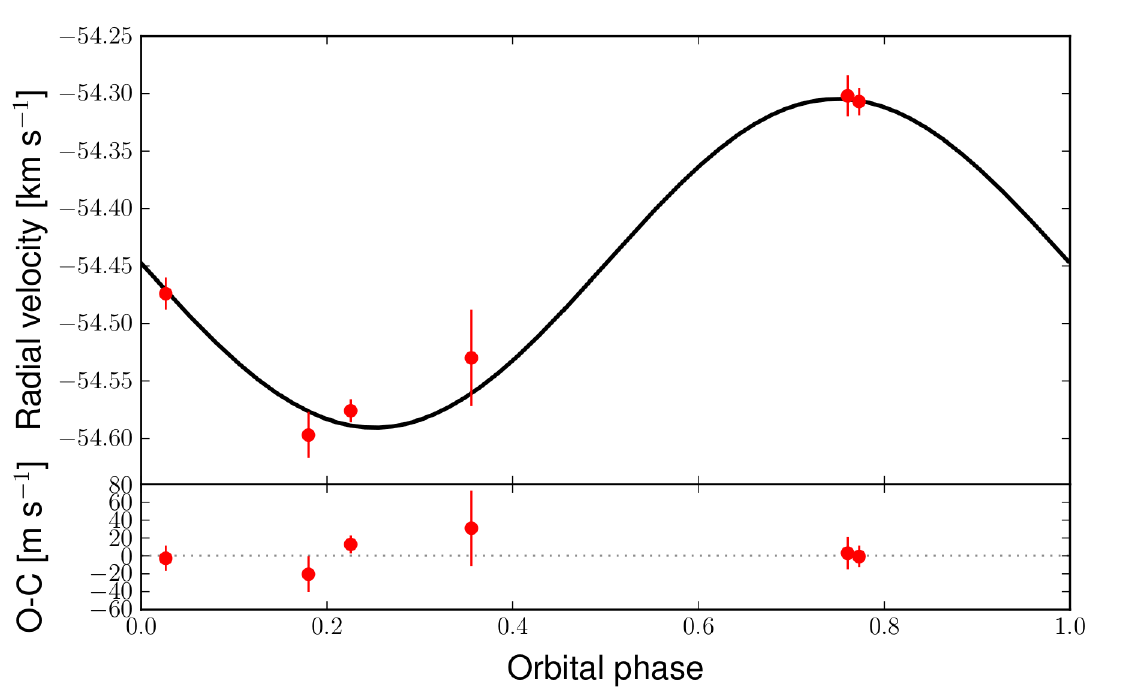}
   \includegraphics[height=6cm,width=9cm]{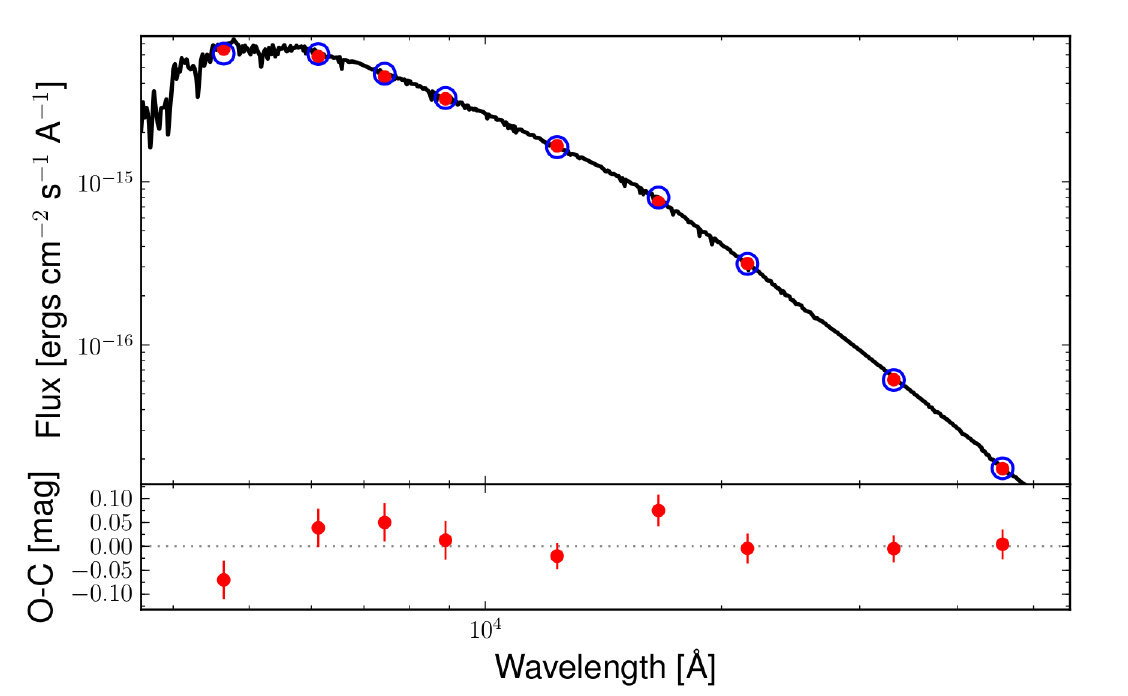}
   \caption{{\sl top panel:} The best-fit model overplotted as a solid line to the phase-folded transit in the phase space. Black dots are long cadence individual points of Kepler-412b transit. {\sl Middle panel:}  The phase-folded radial velocity measurements performed with \textit{SOPHIE}.  {\sl Bottom panel:} The flux integrated in each of the photometric band. The residuals are given at the bottom of each plot. }
\label{AllBest}
\end{figure}

For the photometric time-series analysis, we kept the light curves acquired in long cadence but used the short cadence ones to tune the oversampling parameter following the method that \cite{Kipping2011} originally advocated, and which was later further developed by \cite{Southworth2012}. We first derived the $a/R_{\star}$, $R_p/R_{\star}$, and the inclination $i$ from the transits acquired in short cadence. We then did the same exercise on long cadence transits, using different values of the oversampling for the model until we found the same values of the transit parameters than those derived from short cadence. We found seven as the optimal value and  the modeling was performed on long cadence light curves oversampled to this value. 

The light curves from the different quarters were gathered by seasons to account for change in the contamination value and in the noise budget from one season to another. For the quarters of Season 0, we used the contamination value given in the light curve fits header.  For the others seasons, the contamination values were let as free parameters in the fitting process. Indeed, some authors \citep[][e.g.]{Diaz2013KOI205} report that the crowding factors in the KIC, and beyond that the contamination ones, might be erroneous, which could lead to wrong estimates of the transit depth. In the case of Kepler-412, while the dispersion for the contamination values of a given season is small (0.3\% at most), it varies from about 5.5 \% for Seasons 0 and 1, to 9.5\% and 3.8\% for Seasons 2 and 3, respectively. We thus choose not to adopt the \kep\ contamination values, but let them vary.

 To obtain a clean light curve for the light curve modeling, we reproduced the low-frequency variations of the light curve derived from the periodogram (see Sec.~\ref{photom}), using a Savitzky-Golay smoothing filter with a 24-hour window. The resulting smoothed light curve were used to normalized the original light curve. The goal of this procedure was to remove the star's activity signature and uncorrected instrumental effects from the light curve. Finally, outliers were corrected with a three-sigma clipping performed on the light curve folded at the planet's orbit. The resulting light curve was fitted in time to propagate the error on the ephemerids to the rest of parameters.

The normalized phase-folded light curve was modeled using the EBOP code \citep{Popper1981} provided in the JKTEBOP package \citep{Southworth2011}. It takes the mutual eclipses of both members of the system, the tidal ellipsoidal distortion induced by the planet and  the reflected light of both components of the system into account. We also included the Doppler beaming and we used the formalism described in \cite{Bloemen2011}.  The radial velocities were described by an eccentric Keplerian orbit. The model for the SED modeling was calculated by interpolating the PHOENIX/BT-Settl synthetic spectral library \citep{Allard2012}, scaled to the distance and the color index $E(B-V)$ that were set as free parameters. 

The three sets of data were then fitted simultaneously. To account for systematics errors in each data set that could result in an additional noise, we also introduced a jitter value for each data set. In total, the model accounts for 31 free parameters. The parameters of the model were adjusted to the data using a MCMC algorithm with an adaptive step size, as described in detail by \citep{Diaz2013b}.  Thirty-five chains of 200 000 steps were carried out with initial values randomly drawn from the joint priors. For the latter, we used uniform or Jeffreys distributions for the priors but the stellar density, the \teff, and the metallicity, which were described by a normal distribution. In all cases, the width of the distribution was chosen large enough not to bias the posterior.  For the transit parameters, epoch, and period, we used \cite{Batalha2013} values for the priors. For the parameters related to the host star we used the values obtained from the spectroscopic analysis (Sect.~\ref{spectro}). 

In PASTIS, the burn-in interval of each of the  chains is computed by comparing the mean and standard deviation of the last 10\% of the chain to preceding fractions until a significant difference is found. The correlation length is computed for each of the parameters, and the maximum value is retained and used to thin the chain so that the samples in the chain are independent. The non-convergence is verified using the Gelman and Rubin statistics, and only the chains that show no signs of converging are merged into a single chain. In the case of Kepler-412 system, from the initial 35 chains, 32 passed the Gelman and Rubin statistics test and were used to make the final merged chain. The burn-in interval was between $\sim$ 34 to 50 \%, and the correlation lengths for each chain were between $\sim$ 1110 to  6900 points, depending on the chain.
The median values obtained for the fitted parameters of the Kepler-412 system through this modeling approach are given in Table~\ref{AllParams}. This table also gives  derived parameters and the 68.3\% central confidence interval associated to all parameters. Figure~\ref{AllBest} shows the corresponding best-fit model for the transit, the radial velocity measurements, and the SED. We checked the consistency of the various parameters of the best-fit model with those inferred from analyses used as priors or complementary ones. 

We verified whether our fitted quadratic limb-darkening coefficients agree with the theoretical ones, derived from the photospheric parameters (see Sect.~\ref{spectro}). Using the tables of \cite{Claret2011}, the expected limb-darkening coefficients are $u_a = 0.395 \pm 0.015$ and $u_b = 0.2702 \pm 0.0075$. The best values (Table~\ref{AllParams}) agree within the error bars with the predicted ones. In our modeling, the light curve and the RV measurements yield independent estimates of the mass ratio that we went on to compare. The mass ratio derived from the ellipsoidal variations is $q = 0.00078 \pm 0.00014$, a value nicely consistent with the one calculated from the planet's mass derived from RV amplitude and the star's mass, which is $0.000772 \pm 0.000063$. 
   \begin{figure}
   \centering
   \includegraphics[height=6cm,width=9cm]{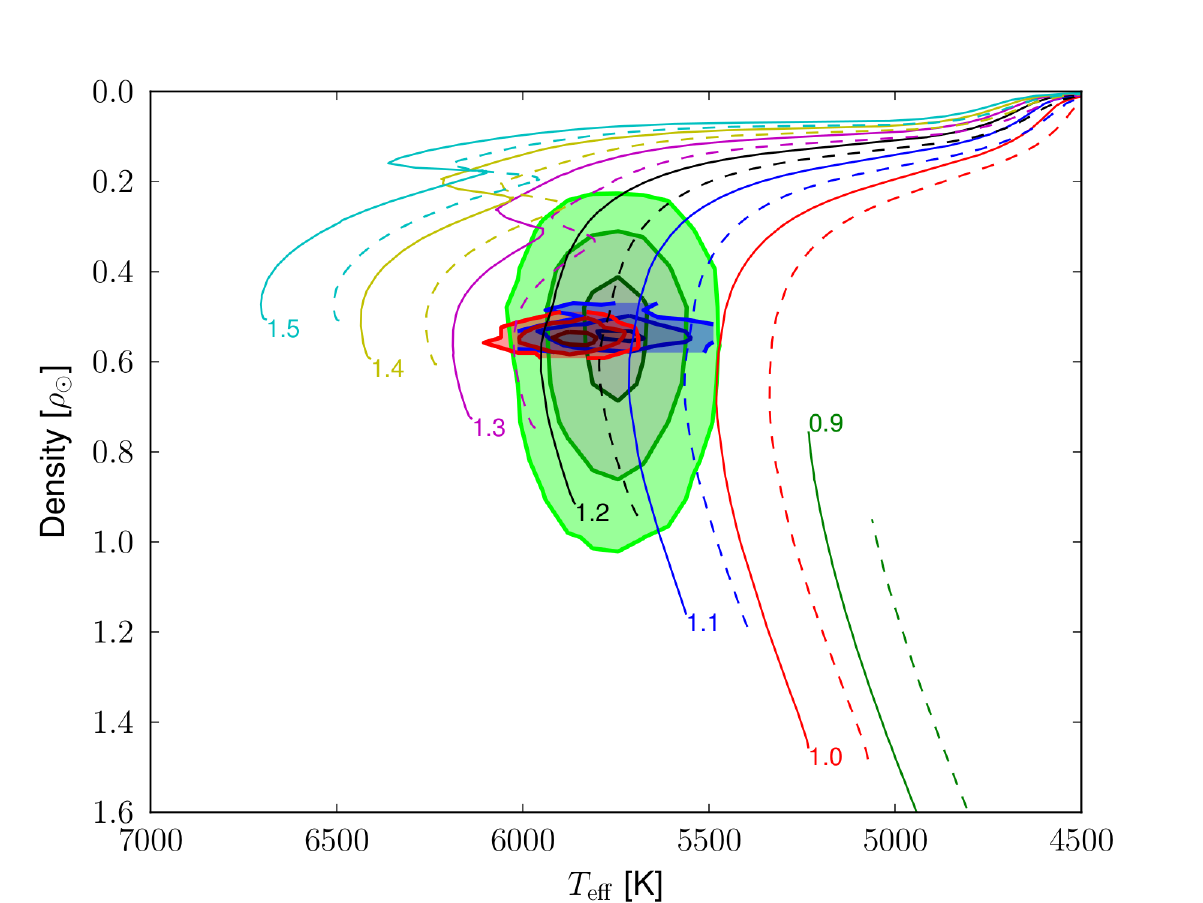}
   \caption{Mean stellar density versus \teff\ diagram. The colored area correspond to the values derived from (i) in green, the spectroscopic analysis, the \logg\ being converted in mean density; (ii) in blue, the atmospheric parameters derived from the spectroscopic analysis combined with the mean stellar density from the transit fitting; (iii) in red, the best model obtained by modeling the SED, the light curve, and the radial velocity measurements. The contours correspond to the 1$\sigma$, 2$\sigma$,  and 3$\sigma$ regions. The curves are STAREVOL stellar evolutionary tracks plotted for a few masses as labeled, and for each mass with two different metallicities: 0.2 (solid line) and 0.3 (dashed line).
  }
\label{stTracks}
\end{figure}

The eccentricity was set as a free parameter. We found a negligible eccentricity  $e < 0.0377$ at 99\% probability interval (0.0062 at 68.3\%). This result is constrained by both the SOPHIE radial velocity measurements and the secondary eclipse. We thus decided on a circular orbit. 

We next compared the stellar parameters derived from the spectroscopic analysis and used as priors (Sect.~\ref{spectro}) to those fitted in PASTIS through the complete modeling of all data sets, that is, \teff = 5866 $\pm$ 60 K and metallicity  $[M/H]$ = 0.20 $\pm$ 0.12. Those values that are within 1$\sigma$ of our spectroscopic values as shown in Fig.~\ref{stTracks}. As usually, the star's mass and radius are better constrained from the transit fitting than by the surface gravity.  From the star's mass and radius estimated by the complete modeling carried out with PASTIS, that is, \Ms\ =  1.17 $\pm$ 0.09 \Msun\ and \Rs\ =  1.28 $\pm$ 0.03 \Rsun, we derived a stellar surface gravity, \logg\ =  4.285 $\pm$ 0.015, in good agreement with the spectroscopic value. 

In conclusion, Kepler-412 system appears to be composed of a solar-type star, slightly enriched in metals, and of a Jupiter-like planet with \Mp=0.939 $\pm$ 0.085 \MJ\ and \Rp = 1.325 $\pm$ 0.043 \RJ on an almost circular orbit. 


\section{Secondary eclipse and planetary phase function }
Kepler-412 phase-folded light curve shows a clear phase function and a significant secondary eclipse (Fig.~\ref{OOT}), which were both modeled in the fitting procedure (see Sect.~\ref{modeling}). This offers the opportunity to probe the overall physical properties of the planet's atmosphere. 

From the eclipse depth, one can set constraints on the geometric albedo and the day side temperature. To that purpose, we used the same methodology as presented in \cite{Santerne2011} with the formalism developed by \cite{Cowan2011}, which assumes the planet is tidally locked. This formalism takes $\varepsilon$, the fraction of the energy which is received by the day side of the planet and transferred to its night side into account. We recall that  $\varepsilon = 0$ when the day side re-emits the energy received and when there is no heat transfer to the night side, and $\varepsilon = 1$ when the energy is homogeneously redistributed over the whole planet's surface. While this methodology has already been followed by several authors \citep[e.g.,][]{Mazeh2012,Esteves2013}, for the sake of clarity, we recall the main equations used in this calculation. The equilibrium day side temperature is a function of both the redistribution factor and the Bond albedo $A_B$: 
   \begin{figure}
   \centering
   \includegraphics[height=6cm,width=9cm]{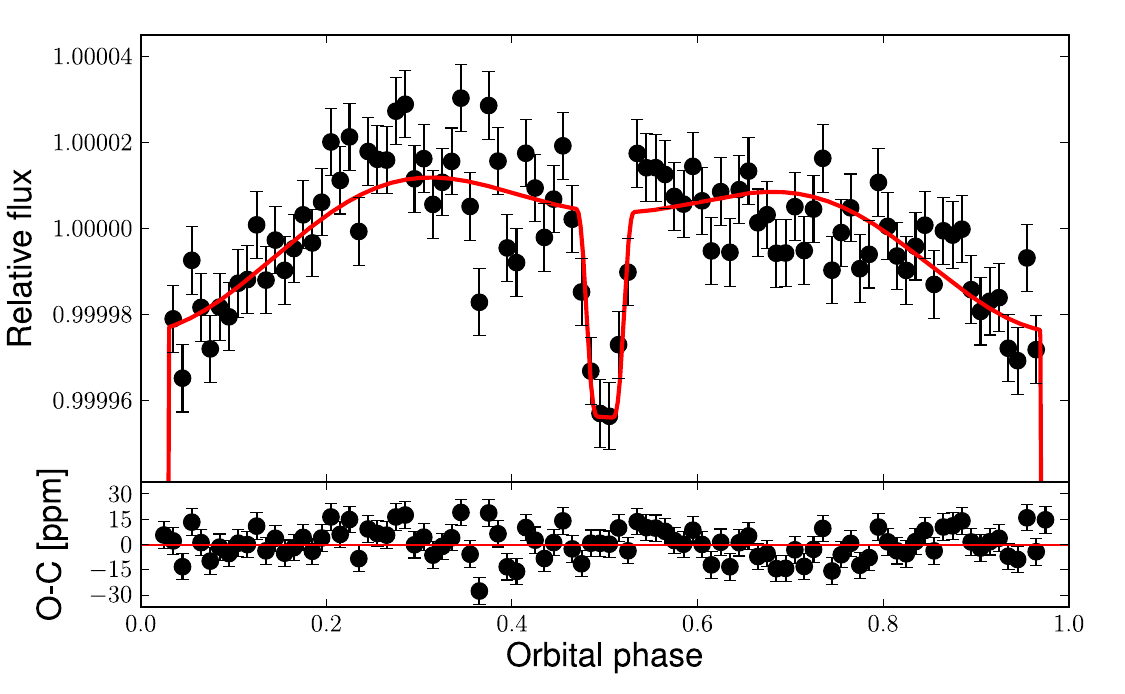}
   \caption{The out-of-transit LC of Kepler-412 in the phase space. The phase bins are 25~min. The best model is overplotted as a full line, and the residuals are shown in the bottom panel.}
\label{OOT}
    \end{figure}

\begin{equation}
T_{eq,D} = T_{eff,\star} \displaystyle{ \left(R_\star \over a \right)^{1/2}  } \left[ \left( \displaystyle{2 \over 3} -  \displaystyle{5 \over 12} \varepsilon \right)~ \left( 1 - A_B \right)  \right]^{1/4}, 
 \end{equation} 
and the night side temperature is given as:
\begin{equation}
 T_{eq,N} = T_{eff,\star} \displaystyle{ \left(R_\star \over a \right)^{1/2}  } \left[ \left( \displaystyle{\varepsilon \over 4} \right)~ \left( 1 - A_B \right)  \right]^{1/4}.
 \end{equation}

For the two extreme cases, $\varepsilon$ = 0 and $\varepsilon$ = 1, of a non reflective planet (A$_B$ = 0), we calculated the range in equilibrium temperature of the  planet's day side. We found the equilibrium temperature of the day side varies from 1828~K in the case of a  complete redistribution to 2336~K when there is a complete re-emission from the day side. 
From the measured eclipse depth and assuming the planet radiates as a black body, we then estimated the planet's brightness temperature in the \kep\ bandpass:
\begin{equation}
F_{ecl}  =   \displaystyle{\left(R_p \over R_\star \right)^2} \displaystyle{ \int{ B_\lambda (T_{B,D}) T_K d\lambda} \over {\int {T_K F_\lambda d\lambda}}} + A_g \displaystyle{ \left(R_p \over a \right)^2}.
 \end{equation} 

\begin{figure}[ht]
   \centering
   \includegraphics[height=6cm,width=9cm]{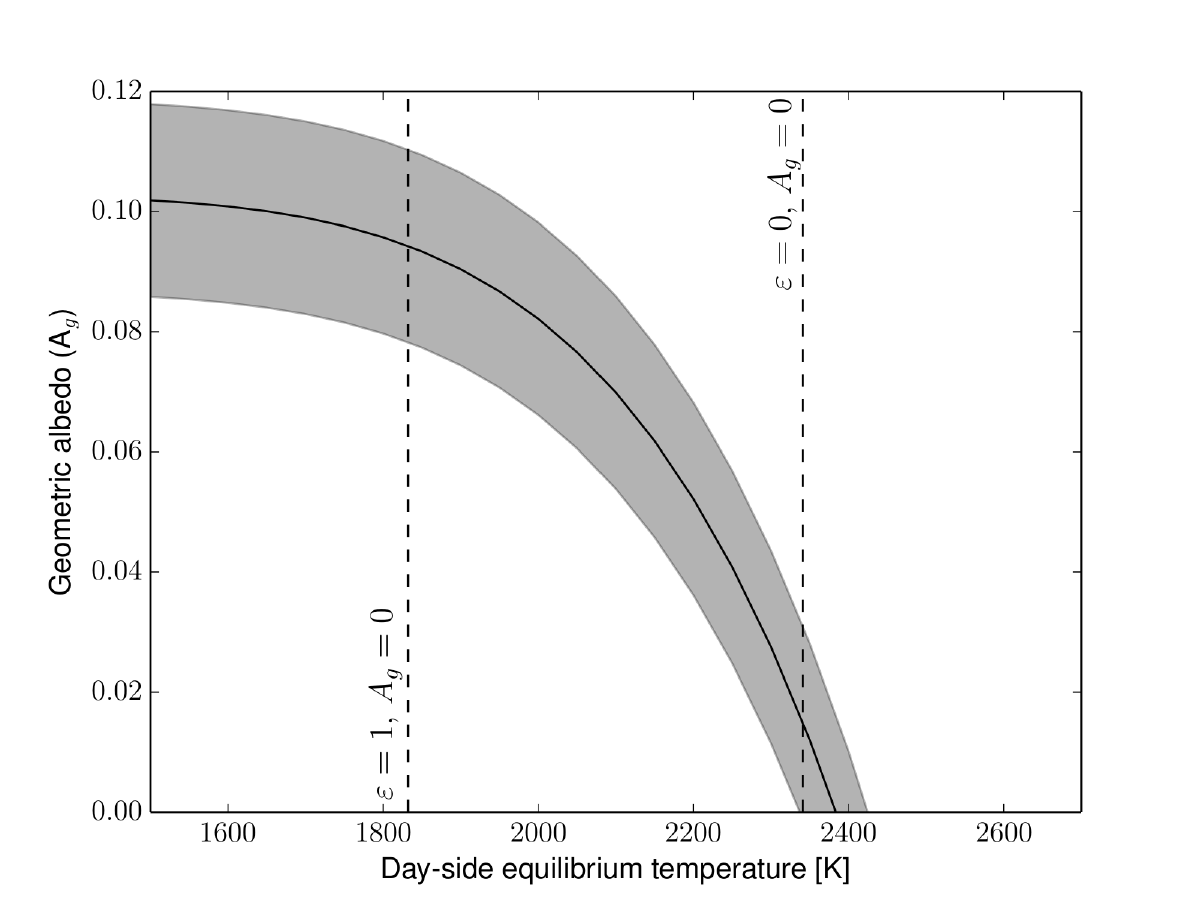}
   \caption{ Geometric albedo as a function of the planet day side brightness temperature from equation (3). The vertical lines correspond to the extreme values for a planet that absorbs all the incoming flux:  $\varepsilon = 0$ when all the energy received on the day side is reemitted;  $\varepsilon = 1$ when all the energy received on the day side is fully re-distributed.The gray band indicates the albedo values allowed by the 1$\sigma$ uncertainty on the occultation depth, which is assumed to be the dominant one.}
\label{TdAlbedo}
\end{figure}

To that purpose, we used the synthetic spectrum of a solar-like star calculated from Kurucz's models \citep{Kurucz1993} to calculate the stellar flux, and we integrated it in the \kep\ bandpass. In the 
unrealistic case where the thermal emission is solely responsible for the secondary eclipse, we found a brightness temperature of the day side of 2380 $\pm$ 40~K.
We then calculated the planet's brightness temperature in the Kepler bandpass (Fig.~\ref{TdAlbedo}), in the more realistic case where both the reflected light and the thermal emission contribute to the eclipse depth.

We found that, in the range defined by the equilibrium temperatures, the geometrical albedo is between 0.013  $^{+0.017}_{-0.013}$ when all the energy received by the planet day side is re-radiated, and  0.094 $\pm$ 0.015  in the extreme case where the eclipse depth in the \kep\ bandpass is due to the reflected energy alone. This value is an upper limit since, even in the visible, one indeed expects the thermal component of the planet's emission to contribute to the phase function. 

\begin{figure}[ht]
   \centering
   \includegraphics[height=6cm,width=9cm]{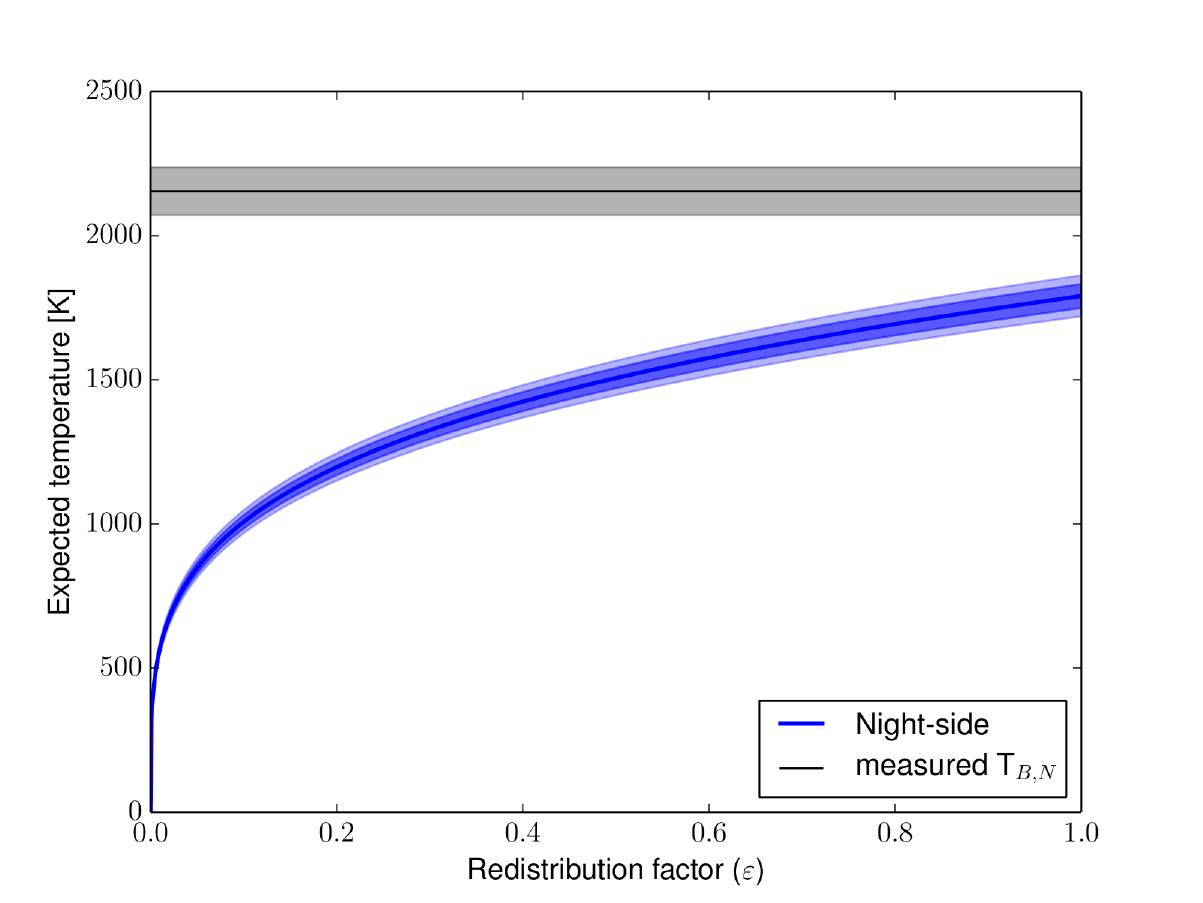}
   \caption{The expected equilibrium temperature of the night side (blue)  as a function of redistribution factor from eqn. 1.
   The dark band gives the limits that correspond to the range of values for the geometrical albedo. The lighter band indicates 
   the 1$\sigma$ uncertainties of each curve, including the uncertainties in the star's temperature and in $R_\star/a$. 
The night side temperature and its associated uncertainty estimated from the night flux is plotted in gray. }
\label{TDistrib}
\end{figure}

The phase function exhibits a significant amplitude that provides information on the night side of the planet. From the difference between the depth of the eclipse and the amplitude of the phase function, we derived a significant night side flux F$_n$ = 18.7 $\pm$ 7.3 ppm.  
Using this value, assuming a black body emission from the planet, and using PHOENIX/BT-Settl synthetic spectra for the star, we found the planet night side brightness emission of T$_{B,N}$ = 2154 $\pm$ 83~K in the \kep\ bandpass. In principle, this might allow evaluation of the redistribution factor between the two planet's faces. 
Figure~\ref{TDistrib} shows how this night side temperature compares with the expected equilibrium night side temperature (from eqn. 2) as a function of the redistribution factor, using the range of values we obtained for the geometrical albedo.   
The measured night side temperature is clearly greater than one would expect even in the case of a planet with a homogeneous heat distribution. There is indeed no overlap between the two curves for the night temperature. A similar situation has been  recently reported in the case of TrES-2 \citep{Esteves2013}. As suggested by these authors, this discrepancy might be related to the physical properties of the planet's emitting layers probed in the \kep\ bandpass, their temperature being higher than the equilibrium temperature. Another possibility would be that an additional source of heating in the planet interior, not included in the simple model we used, might be responsible for this high temperature of the night side of the planet.  

We noted that the phase light curve in Figure~\ref{OOT} might exhibit an eastward asymmetry. However, we found that this suspected asymmetry is not significant and did not try to assess it to atmospheric circulation properties.

\begin{table}[ht]
\begin{minipage}[!]{8.0cm}  
\caption{\label{second}  Fitted and derived parameters for the secondary and the phase curve. }
\renewcommand{\footnoterule}{}     
\begin{tabular}{lll}
\hline
\hline
 F$_{ecl}$ & Depth of secondary eclipse   [ppm]  & 47.4 $\pm$ 7.4 \\  
 A$_{ellip}^a$    & Amplitude of the ellipsoidal variation  [ppm] &  10.2 $\pm$  2.5 \\
 A$_{refl}^a$    & Amplitude of the reflected light  [ppm]  & 14.2 $\pm$ 1.6 \\
 A$_{beam}^a$    & Amplitude of the beaming effect  [ppm] & 1.81 $\pm$ 0.14 \\
 A$_{p}$    & Amplitude of the phase function [ppm] &  28.8 $\pm$ 3.2 \\
 F$_{n}$ & Night side flux  [ppm]  &  18.7 $\pm$  7.3 \\  
\hline
\multicolumn{3}{l}{Derived day side parameters} \\
\hline
A$_{g,max}$    & Geometric albedo & 0.094 $\pm$ 0.015 \\
A$_{g,min}$    & Geometric albedo &  0.013  $^{+0.017}_{-0.013}$ \\
T$_{eq,min}$    & Equilibrium temperature ($\varepsilon$ = 1) [K] & 1828 $\pm$ 22  \\
T$_{eq,max}$    & Equilibrium temperature  ($\varepsilon$ = 0) [K]  & 2336 $\pm$ 27 \\
T$_{B,D}$    & Brightness day side temperature [K] & 2380 $\pm$ 40 \\
\hline
\multicolumn{3}{l}{Derived night side parameter} \\
\hline
T$_{B,N}$    & Night side temperature [K] &  2154 $\pm$  83 \\
\hline
\vspace{-0.5cm}
\end{tabular}
\footnotetext[1]{Calculated using \cite{Mazeh2010} formalism.}
\end{minipage}
\end{table}

\section{Kepler-412b internal structure}
\begin{figure}
   \centering
   \includegraphics[height=7cm,width=9cm]{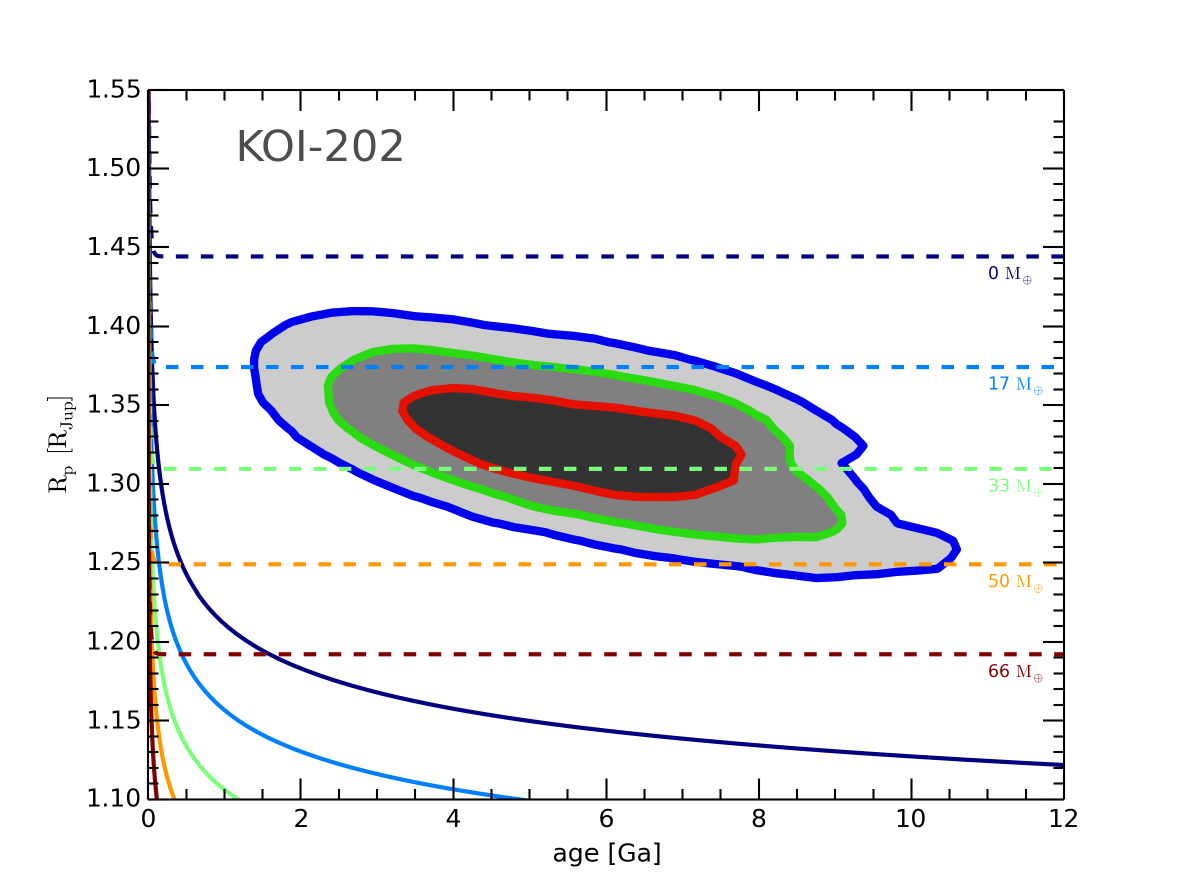}
   \caption{ Evolution of Kepler-412b's radius as a function of the age. The 68.3\%, 95.5\% and 99.7\% confidence regions are denoted by black,
dark gray, and light gray areas respectively. The curves represent the thermal evolution of a 0.939\,M$_{\rm Jup}$ planet with an equilibrium temperature of
1851\,K. Text labels indicate the amount of heavy elements in the planet (its core mass in Earth masses). Dashed lines represent planetary evolution models for which 0.25\% of the incoming stellar flux (about $1.789\times 10^{27}\rm\, erg/s$) is dissipated into the core of the planet, whereas full lines do not account for this dissipation (standard
models).}
\label{intern}
\end{figure}

With a mass of 0.939 $\rm M_{Jup}$, a radius of 1.325 $\rm R_{Jup}$, and an inferred density of just 0.50 $\pm$ 0.05 $\rm g\, cm^3$, Kepler-412b belongs to the inflated class of the hot Jupiter population. Combined stellar \citep[CESAM][]{Morel2008} and planetary \citep[CEPAM][]{Guillot1995, Guillot2010} evolution models of the Kepler-412 system were
calculated with SET \citep[see][]{GH2011, Havel2011}. We used a MCMC algorithm to obtain posterior probability distributions
of the bulk composition of the planet. The results are presented in terms of planetary radii as a function of age in Fig.~\ref{intern}.

Planetary evolution models are calculated in two cases: a ``standard'' case where the planet is assumed to be made of a central rocky core and a solar-composition 
envelope (Fig.~\ref{intern}); and a ``dissipated-energy'' model in which, in addition to the standard case, a fraction (0.25\%) of the incoming
stellar light is assumed to be converted into kinetic energy and then dissipated at the center of the planet \citep[see][for more details]{Almenara2013}.

Only models with dissipation provide solutions for the planetary radius that match the available constraints, resulting in a core mass of
$31\pm 11\rm\,M_{\oplus}$, which translates into a heavy element mass fraction of $0.11\pm 0.04$\footnote{These numbers come from independent MCMC 1-D distributions}. However, we do not know precisely how much the planet could actually dissipate in its interior \citep[e.g.][]{GS2002, BSB11, Laughlin2011}: reducing by two (respectively 5 and 10) the amount of energy dissipated in the planet's interior, moves the $0 \,M_{\oplus}$ line from 1.444 to 1.391\RJ\ (resp. 1.332 and 1.294)  (Fig.~\ref{intern}), which corresponds to a 3.7\% reduction in radius (resp. 7.7\% and 10\%). It is therefore possible that Kepler-412b has no core at all. On the other hand, the high metallicity of the star tends to indicate that the planet should probably have a certain amount of heavy elements \citep{Guillot2006}. Whether they are dispersed in the envelope \citep{BCB08} or are all condensed into the core is not clear, but does not significantly affect our results.

\section{Kepler-412 system in context }

At visible wavelengths, phase curve and secondary eclipse have been measured for only a dozen validated planetary systems only (Table~\ref{AllPhase}). Two are CoRoT planets, but the large majority of the sample are systems found from \kep\ light curves. All these planets are close-in giants that cover a range of masses over nearly one order of magnitude, from Kepler-12b whose mass is only 0.43\MJ\ to KOI-13b, which has a mass greater than 8 \MJ. All of them also present a measured geometric albedo that is much lower than giant planets in the Solar System. Their low albedo is in agreement with models of irradiated planets \citep{Fortney2006} but here again with noticeable diversity. 
  
In an attempt to explain the diversity in albedo of the close-in population and the lack of any clear correlation between the albedo and the incident stellar flux,  \cite{Heng2013} explored a possible link between the albedo and the atmospheric properties through a combination of the variation in size of the cloud particles and atmospheric circulation. From analytical models, they suggest some possible interpretation in terms of atmospheric characteristics for a  few configurations. In that context, with its low albedo but clear sinusoidal optical phase curve, Kepler-412b somehow appears in between the rough classes proposed by these authors that connect the optical phase curve to  the atmospheric properties based on size of cloud particles, showing again the complexity of the atmosphere properties. 

\begin{figure}
   \centering
   \includegraphics[height=6cm,width=8cm]{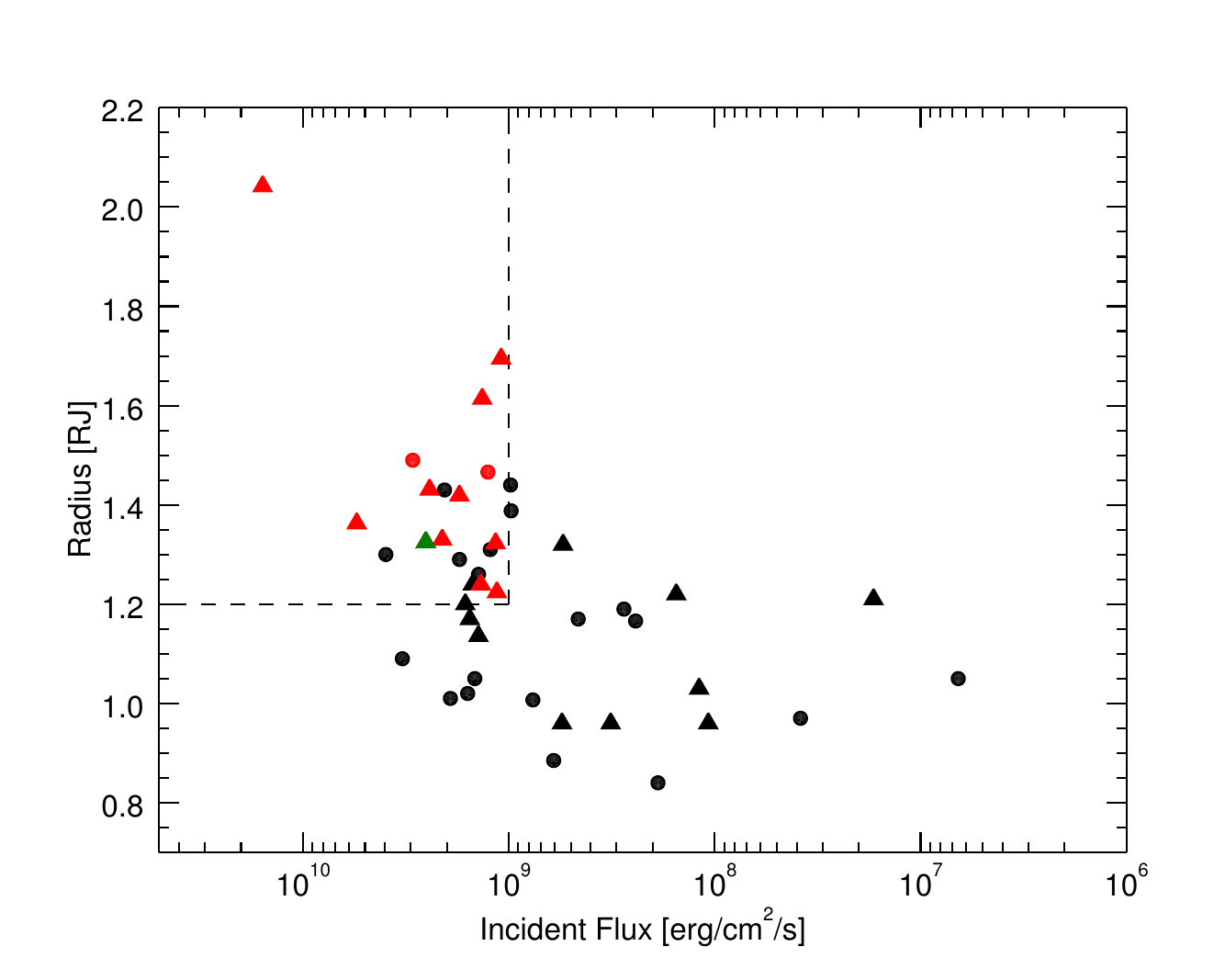}
   \caption{Radius versus the total incident stellar flux for giant planets observed from space. Triangles are \kep\ planets, disks are CoRoT ones. The red symbols correspond to those from both missions with phase function and/or secondary detection at visible wavelengths. Kepler-412b is highlighted by a green triangle.}
\label{secplot}
\end{figure}

Figure~\ref{secplot} shows how the giant planets with optical secondary eclipse or phase function reported in the literature compare to other transiting giant planets from space-based surveys in the incident flux - planetary radius diagram. We excluded all of the many giants detected from ground-based surveys since the expected secondary signature at visible wavelengths is beyond the reach of existing ground-based facilities, with the exception of the bright hot Jupiter HD189733b for which evidence of scattering polarization in the planetary atmosphere has been reported \citep{Berdyugina2011}. On this plot, giant planets with detected albedo gather in a squared region roughly above 1.0 10$^{9}$  erg s$^{-1}$ cm$^{2}$ and 1.2 \RJ. They seem to share the characteristic of having an inflated radius. In that picture, Kepler-41b might have appeared as an outlier \citep{Santerne2011}. However, \cite{Bonomo2013} have recently revised the planet's parameters thanks to analyzing \kep\ short cadence quarters combined with new spectroscopic data of the host star. From their study, they updated the planet's radius to 1.24 \RJ\ that makes Kepler-41b join the group of inflated planets. We noted that planets that lie in this region of the plot with no eclipse detection are CoRoT planets. \cite{Parviainen2013} carried out a statistical study whose aim was to identify secondary eclipses of CoRoT giant planets from numbers 1 to 23. They did not report any clear new detection but statistically significant eclipses for two planets, CoRoT-6b and CoRoT-11b, and three systems with marginal detection, CoRoT-13b, CoRoT-18b, and CoRoT-21b. CoRoT-11b, 18b, and 21b are indeed in the region delimited by systems with detected secondary. If the trend on Fig.~\ref{secplot} is real, it would indicate that for faint targets the detection of secondary from CoRoT light curves requires a precision beyond the instrument's capacity. 

Even though the nature of the mechanisms responsible for radius inflation for the close-in giants is still being debated, evidence accumulates of a correlation between the inflation and the intensity of the stellar incident flux \citep{Fortney2008,DemorySeager2011}. Figure~\ref{secplot} appears to argue for an explanation based on one of the mechanisms responsible for the radius anomaly to also be responsible for the atmosphere reflectivity of this pM-class planets through the influence on the atmospheric circulation or vertical structure. 
However, with such a limited sample and without the support of atmosphere models of irradiated planet, it is difficult to draw firm conclusions. Definitely, this tendency would require a much larger sample to be confirmed. 

\section{Summary}
Radial velocity measurements carried out with the SOPHIE high-precision spectrograph at the OHP allowed us to secure the planetary nature of Kepler-412b, a planet candidate whose transits were detected in the \kep\ light curves. We also derived its fundamental parameters by performing the combined analysis of the radial velocity measurements, the complete series of \kep\ quarters (Q1 to Q15), and the host star properties, through a Bayesian approach implemented in the PASTIS software. We found that with a mass of 0.939\MJ\  and a radius of 1.325\RJ, the planet is a new member of the hot Jupiter population with an inflated radius. For this low-density planet, we found that models with dissipation are required in order to reproduced the planet's radius at its current age. The model suggests there is a significant amount of heavy elements in the interior of the planet, maybe in the form of a massive core of about 31\ME, with an extended envelope of H-He.

From the phase function and the secondary eclipse, we derived a very low geometric albedo less than 0.1, and a night side temperature much higher than the maximum hemisphere-averaged expected equilibrium one. 
 The \kep\ data alone do not give us strong leverage on the temperature structure of the planet's atmosphere and do not allow us to investigate it in detail as it has been done for some of the \kep\ planets with additional measurements of the eclipse in the near infrared. However, the comparison to other close-in giants shows a possible correlation between the secondary detection and one of the mechanisms that might be responsible for the radius anomaly.  This needs to be validated with a much larger sample, but 
if confirmed, it could suggest that the differences in atmospheric properties of close-in giants with a marked inflated radius and the non-inflated ones is also reflected in their phase light curve properties.

\begin{table*}[ht]
\caption{\label{AllPhase}  Planetary systems with phase curve and/or secondary detected.}
\renewcommand{\footnoterule}{}     
\begin{tabular}{llllllll}
\hline
\hline
System              & Teff  &  a             & T$_{eq}$        & M$_p$ & R$_p$ & P    \\
\hline
Kepler-5$^{1,2,3}$   & 6297 $\pm$ 60 &  0.05064$\pm$ 0.0018 &   1752 $\pm$ 17      & 2.112  & 1.431 & 3.5484  \\
Kepler-6$^{4,2,3}$   & 5647 $\pm$ 44 &  0.04561 $\pm$ 0.0006  &   1451 $\pm$ 16      & 0.668 & 1.323 & 3.2347  \\
Kepler-7$^{5,3}$      & 5933 $\pm$ 44   &  0.06246 $\pm$ 0.00046    &   1586 $\pm$ 13    & 1.443 & 1.614  & 4.8854 \\
Kepler-8$^{6,2,3}$   & 6210 $\pm$ 150  &  0.04833 $\pm$ 0.002 &   1638 $\pm$ 40      & 0.583   & 1.419 & 3.5225 \\
Kepler-12$^{7,3}$    & 5947 $\pm$ 100 &  0.05563 $\pm$ 0.007  &    1477 $\pm$ 26     & 0.432   & 1.695 & 4.438   \\
Kepler-17$^{8,9,3}$ & 5781 $\pm$  85 &  0.0268 $\pm$ 0.0005  &    1655 $\pm$ 40  & 2.477   & 1.330   & 1.4857 \\
Kepler-41$^{10,11,3}$ & 5620 $\pm$ 140 &  0.0303 $\pm$ 0.0009   &      1567 $\pm$  41  & 0.540  & 1.240   & 1.8555 \\
KOI-13$^{12,13,2}$ & 8500 $\pm$ 400 &   0.0514 $\pm$ 0.0020  &   2885 $\pm$  140 & 8.500     & 2.042 & 3.5225 \\
HAT-P-7$^{14,2,3}$ & 6350 $\pm$ 80 &  0.0355 $\pm$ 0.0044 &  2139 $\pm$ 27          & 1.779 & 1.363   & 2.204    \\
TRES-2$^{15,2,3}$  & 5850 $\pm$ 50 &  0.0367$\pm$ 0.0013 &  1444 $\pm$ 13    & 1.193 & 1.224 & 2.470     \\
CoRoT-1$^{16}$      & 5950 $\pm$ 150 &  0.0254 $\pm$ 0.0004 &   1896 $\pm$  52  & 1.037 & 1.490 & 1.5089   \\
CoRoT-2$^{17,18}$  & 5625 $\pm$ 120 &  0.0281 $\pm$ 0.0004 &   1536 $\pm$ 35  & 3.273   & 1.466 & 1.742    \\
\hline
\end{tabular}
\tablefoot{ \tablefoottext{1}{\cite{Koch2010}},\tablefoottext{2}{\cite{Esteves2013}},\tablefoottext{3}{\cite{Heng2013}},\tablefoottext{4}{\cite{Dunham2010}}, \tablefoottext{5}{\cite{Demory2011}}, \tablefoottext{6}{\cite{Jenkins2010}}, \tablefoottext{7}{\cite{Fortney2011K12}}, 
\tablefoottext{8}{\cite{Desert2011}},\tablefoottext{9}{\cite{Bonomo2012}},
\tablefoottext{10}{\cite{Santerne2011}}, \tablefoottext{11}{Bonomo et al., {\sl to be submitted}}, \tablefoottext{12}{\cite{Szabo2011}},\tablefoottext{13}{\cite{Mazeh2012}} ,
\tablefoottext{14}{\cite{Christiansen2010}}, \tablefoottext{15}{\cite{Barclay2012}}, \tablefoottext{16}{\cite{Alonso2009a}}, \tablefoottext{17}{\cite{Snellen2009}, \tablefoottext{18}\cite{Alonso2009b}}. In some cases, asymmetric error bars have been symmetrized.
}
\end{table*}

\begin{acknowledgements}
This work is based on observations collected with the NASA's satellite Kepler, the SOPHIE spectrograph on the 1.93-m telescope at Observatoire de Haute-Provence (CNRS), France. The authors thank the staff at Haute-Provence Observatory. They also acknowledge the PNP of CNRS/INSU and the French ANR for their support. The team at LAM acknowledges support by grants  98761 (SCCB), 251091 (JMA), and 426808 (CD). RFD was supported by CNES via its postdoctoral fellowship program.
AS acknowledges the support of the European Research Council/European Community under the FP7 through
Starting Grant agreement number 239953. MH was supported by an appointment to the NASA Postdoctoral Program at the Ames Research Center, administered by Oak  Ridge Associated Universities through a contract with NASA. ASB gratefully acknowledges support through INAF/HARPS-N fellowship.

\end{acknowledgements}

\bibliographystyle{aa.bst}
\bibliography{Kepler-412-bib}
\clearpage 
\begin{table}[ht]
\caption{ \label{AllParams} Kepler-412 system: planet and star parameters.}            
\begin{minipage}[!]{17.0cm}  
\renewcommand{\footnoterule}{}   
\begin{tabular}{ll}        
\hline\hline                 
\multicolumn{2}{l}{\emph{Ephemeris and orbital parameters: }} \\
\hline
Planet orbital period $P$ [days] & 1.720861232 $\pm$ 4.7 10$^{-8}$ \\
Primary transit epoch $T_{tr}$ [BJD-2400000] &  54 966.021033  $\pm$ 2.3 10$^{-5}$ \\
Orbital eccentricity $e$  &  0.0038 $\pm$ $_{-0.0032}^{+0.0087}$\\
Argument of periastron $\omega$ [deg] & 125 $_{-49}^{+150}$ \\ 
Orbital inclination $i$ [deg] & 80.89 $\pm$ 0.20  \\
Primary impact parameter\tablefootmark{a} $b_{prim}$ & 0.781 $_{-0.010}^{+0.006}$ \\
Primary transit duration $T_{14}$ [h] & 2.077 $\pm$ 0.015  \\
\hline
\multicolumn{2}{l}{\emph{Fitted transit-related parameters: }} \\
Scaled semi-major axis $a/R_{\star}$ & 4.947 $\pm$ 0.056   \\
Radius ratio $R_{p}/R_{\star}$ & 0.1058 $\pm$ 0.0023 \\
Quadratic limb darkening coefficient $u_a$ & 0.18 $\pm$ 0.44 \\
Quadratic limb darkening coefficient $u_b$ & 0.51 $\pm$ 0.58 \\
$R_p / a $  & 0.02145 $\pm$ 0.00052  \\
\hline
\multicolumn{2}{l}{\emph{Data-related parameters: }} \\
Kepler season 0 contamination  [\%] &  5.47        \\
Kepler season 1 contamination  [\%]  & 5.24 $\pm$ 0.15 \\
Kepler season 2 contamination  [\%]  & 3.42 $\pm$ 0.17 \\
Kepler season 3 contamination  [\%]  & 5.69 $\pm$ 0.13 \\
Kepler season 0 jitter  [ppm] & 96.3 $\pm$ 3.4 \\
Kepler season 1 jitter [ppm]  & 117.1 $\pm$ 2.6 \\
Kepler season 2 jitter [ppm]  & 114.5 $\pm$ 4.0 \\
Kepler season 3 jitter [ppm]  &  90.6 $\pm$ 3.1 \\
SOPHIE jitter [\ms] & 12.4$_{-9.9}^{+16.0}$ \\
SED jitter [mags] & 0.028 $\pm$ 0.022 \\
\hline 
\multicolumn{2}{l}{\emph{Fitted RV-related parameters: }} \\
Radial velocity semi-amplitude $K$ [\kms] & 0.142 $\pm$ 0.011 \\
\hline
\multicolumn{2}{l}{\emph{Spectroscopic parameters: }} \\
Effective temperature $T_{eff}$[K] & 5750 $\pm$ 90 \\
Surface gravity log\,$g$ [dex]& 4.30 $\pm$ 0.07  \\
Metallicity $[\rm{Fe/H}]$ [dex]& 0.27 $\pm$ 0.12\\
Stellar rotational velocity {\vsini} [\kms] & 5.0 $\pm$ 1.0	  \\
Stellar microturbulence {\vmic} [\kms] &  1.0 $\pm$ 0.1   \\
Stellar macroturbulence {\vmac} [\kms\ &  2.86 $\pm$1.0 \\
Spectral type & G3 V \\
\hline
\multicolumn{2}{l}{\emph{Stellar physical parameters from combined analysis:}} \\
\mtier\ [solar units] & 0.8184 $\pm$ 0.0092 \\ 
Stellar density $\rho_{\star}$ [solar units]  & 0.547 $\pm$ 0.018 \\
Star mass [\Msun] &  1.167 $\pm$ 0.091 \\
Star radius [\Rsun] &  1.287 $\pm$ 0.035  \\
Stellar rotation period $P_{rot}$ [days]  &  17.2 $\pm$ 1.6 \\
Age of the star $t$ [Gyr] & 5.1 $\pm$ 1.7 \\
Distance of the system [kpc] & 1.056 $\pm$ 0.036  \\
Interstellar extinction $E(B - V)$ &  0.013 $\pm$ 0.015\\
\hline    
\multicolumn{2}{l}{\emph{Planetary physical parameters from combined analysis:}} \\
Orbital semi-major axis $a$ [AU] & 0.02959 $\pm$ 0.00078  \\
Planet mass $M_{p}$ [M$_J$ ]$^b$ &  0.939 $\pm$ 0.085 \\
Planet radius $R_{p}$[R$_J$]$^b$  &  1.325 $\pm$ 0.043  \\
Planet density $\rho_{p}$ [$g\;cm^{-3}$] & 0.501  $\pm$ 0.051\\
\hline       
\vspace{-0.5cm}
\footnotetext[1]{ $b=\frac{a \cdot \cos{i}}{R_{*}} \cdot \frac{1-e^{2}}{1+e \cdot \sin{\omega}}$}
\footnotetext[2]{Radius and mass of Jupiter taken as 71492 km and 1.8986$\times$10$^{30}$ g, respectively.}
\end{tabular}
\end{minipage}
\end{table}
\end{document}